\documentclass[preprint,preprintnumbers,showpacs,aps,amssymb]{revtex4}

\usepackage{graphicx}
\usepackage{bm}
\usepackage{amsmath}

\def\calO{{\cal O}}
\def\calL{{\cal L}}

\def\Bbar{{\bar B}}

\def\hbar{{\bar h}}

\def\qbar{{\bar q}}

\def\ubar{{\bar u}}
\def\vbar{{\bar v}}

\def\kslash{k\hspace{-1.8mm}/}
\def\nslash{n\hspace{-2.2mm}/}

\def\pslash{p\hspace{-1.8mm}/}
\def\vslash{v\hspace{-1.8mm}/}

\def\nn{\nonumber}

\begin{document}
\title{$B$ to light meson form factors with end-point cutoff}
\author{Jong-Phil Lee}
\email{jplee@kias.re.kr}
\affiliation{Korea Institute for Advanced Study, Seoul 130-722, Korea}
\preprint{KIAS-P07010}

\begin{abstract}
Decay form factors of $B$ to light pseudoscalar ($P$) and vector ($V$) mesons 
are analyzed with the momentum cutoff near end point.
The cutoff is caused by possible Cherenkov gluon radiation when an energetic
light parton from the weak vertex travels through the ''brown muck'' of light
degrees of freedom.
The end-point singularities and the double countings are naturally absent.
The soft-overlap contributions where the partonic momenta configuration is 
highly asymmetric are very suppressed in this framework.
A simple calculation gives a plausible results for $B\to P,V$ form factors.
\end{abstract}
\pacs{12.38.Lg, 13.20.He}

\maketitle
\section{Introduction}
Heavy-to-light decays are very important processes in flavor physics.
The successful running of $B$ factories BABAR and BELLE has enabled the 
semileptonic $B$ decays like $B\to\pi(\rho)\ell\nu$ to work out the information
about the least known Cabibbo-Kobayashi-Maskawa (CKM) matrix element $|V_{ub}|$,
for which $B\to\pi(\rho)$ form factors play the crucial roles.
The same form factors enter the nonleptonic modes such as $B\to\pi\pi$,
$B\to\pi\rho$, which are interesting because of the factorization as well as
the angles of CKM unitarity triangle.
Heavy-to-light decay is also responsible for $b\to s$ transitions 
$B\to K^*\gamma$ or $B\to K^*\ell\ell$, which are good probes for new physics 
beyond the standard model(SM).
\par
However, the theoretical understanding of heavy-to-light transition is quite
poor and still controversial.
The transition matrix elements are usually parametrized by the form factors.
They are nonperturbative objects which cannot be calculated from first 
principles, and all the essential features of the decay processes are encoded
on them.
At large recoil limit where the final state meson energy is sufficiently large,
these form factors are not independent.
Conventionally, there are 3(7) decay form factors for $B$ to $P(V)$ decays,
where $P(V)$ stands for the pseudoscalar (vector) mesons.
The number of independent form factors reduces to 1(2) for $B\to P(V)$ at large
recoil limit \cite{Charles:1998dr}.
This is due to the so called spin-symmetry relations.
\par
It is a common lore that there are two kinematically distinctive contributions
to the form factors.
One is the ''soft overlap (or Feynman mechanism)'' where one of the partons in 
the daughter meson carries almost all the momentum.
The other is the ''hard scattering'' where none of the partons in the daughter
meson is in the end-point region of momentum configuration by exchanging
hard gluons.
But there is no general agreement on which of the two is the dominant one.
\par
One of the main issue related is the end-point singularity.
Contributions of hard gluon exchange with the spectator quarks are described
by the convolution integrals involving meson distribution amplitudes (DA) and
some kernel.
At the heavy quark and large recoil limit, the kernel behaves like 
$\sim 1/x^2$ where $x$ is some momentum fraction of the partons while the 
DAs of the final state meson do as $\sim x$ in their asymptotic forms.
The resulting convolution integral diverges at the end point $x\sim 0$,
and this is called the end-point singularity.
\par
The end-point singularity has been dealt in many different ways in different 
theoretical frameworks.
In perturbative QCD (pQCD) 
\cite{Li:1994ck,Keum:2000ph,Keum:2000ms,Kurimoto:2001zj,Li:2001ay},
the end-point singularity is absent due to the Sudakov suppression near the
end-point region.
The point is that when one of the partons carries all the momentum, the
remaining part of the daughter meson cannot shield the color charge of the
fast parton, so many gluons should be emitted \cite{Kurimoto:2001zj}.
Thus the soft overlap contribution is very suppressed in this picture, and 
the hard scattering is the dominant contribution.
But the Sudakov suppression is not generally accepted in the literature.
In Ref.\ \cite{Genon,Lange:2003pk,Hill:2004if}, it is argued that the Sudakov
suppression is not severe at the heavy quark scale $m_B\sim 5.3$ GeV (for
more recent discussions, see \cite{Wei:2002iu}) .
\par
On the other hand, in Ref.\ \cite{Beneke:2000wa}, the problem of end-point
singularity is evaded by absorbing the singular terms into the so called
''soft form factor''.
Here the heavy-light $B\to M$ ($M$ is a light meson) form factors $f_i$ are 
compactly written as
\begin{equation}
f_i(q^2)=C_i\xi_i^M(E)+\phi_B\otimes T_i\otimes\phi_M~,
\label{soft}
\end{equation}
where $\xi_i^M(E)$ are the soft form factors for $M$ of energy $E$, and
$C_i$ are the hard vertex renormalization;
$T_i$ are hard kernels which are convoluted ($\otimes$) with the DAs $\phi_B$
and $\phi_M$, and $q$ is the momentum transfer.
It was shown that the soft form factors $\xi_i^M$ satisfy the spin-symmetry 
relations.
The second term of convolution arises from the hard spectator interactions.
Terms involving end-point singularities are already absorbed into $\xi_i^M$,
so the remaining convolutions are end-point finite.
They are shown to break the spin-symmetry relations.
The soft form factors $\xi_i^M$ are totally nonperturbative objects;
they cannot be calculated from first principles, and they contain the 
end-point singularities.
The meson DAs $\phi_{B,M}$ are also nonperturbative, but they can be treated
systematically and are well understood.
Other quantities $C_i$ and $T_i$ are perturbatively calculable.
In fact, one of the great merit of Eq.\ (\ref{soft}) is the perturbative
calculability of the hard kernel, $T_i$.
This is the main philosophy of the QCD factorization \cite{BBNS}.
From the numerical analysis, the authors of Ref.\ \cite{Beneke:2000wa} found
that the symmetry breaking convolutions contribute about 10\%;
they conclude that heavy-light form factors are largely from the soft form 
factor.
\par
The advent of the soft-collinear effective theory (SCET) \cite{SCET} shed new
lights on the heavy-to-light decays.
In this framework, the $B\to M$ form factors are described as
\cite{Bauer:2002aj,Hill:2004if,Manohar:2006nz}
\begin{equation}
f_i=T^{(i)}(E)\zeta^{BM}(E)
+N_0\phi_B\otimes C_J^{(i)}\otimes J\otimes\phi_M~,
\label{SCET}
\end{equation}
where $T^{(i)}$ and $C_J^{(i)}$ are the hard functions and $J$ is the jet
function, and $N_0=f_Bf_M m_B/(4E^2)$ with $f_{B,M}$ being the meson decay
constants.
Here $\zeta^{BM}$ is the SCET version of the ''soft'' form factor.
Just as in \cite{Beneke:2000wa}, the end-point singular terms are absorbed into
$\zeta^{B\pi}$, and as a whole satisfy the spin-symmetry relations.
From the fact that the same form factors enter the nonleptonic two-body $B$
decays, Ref.\ \cite{Bauer:2004tj} showed that the two contributions of
Eq.\ (\ref{SCET}) are comparable in size.
This point differs from the QCD factorization (QCDF) analysis
\cite{Beneke:2004bn}, where the hard scattering contribution is very small.
\par
Although the brief summary above shows impressive achievements in heavy-to-light
transitions, there are still some ambiguities and confusions.
First of all, the quantity $\zeta^{B\pi}$ in Eq.\ (\ref{SCET}) is not ''soft''
in the sense that the involved quarks are not in the asymmetric momentum
configuration; it is defined by the {\em collinear} quarks.
In this context, the SCET description of Eq.\ (\ref{SCET}) is much closer to
the pQCD prediction where the contributions from the asymmetric momentum
configuration (i.e., soft overlap) are highly suppressed.
\par
It is very helpful to see how the soft overlap is identified in the 
light-cone sum rule (LCSR).
The LCSR has improved their predictions for the $B\to M$ form factors 
\cite{Ball:2001fp,Ball:2004rg}.
At tree level after the Borel transformation, the weak form factor is
proportional to \cite{Ball:1997rj}
\begin{equation}
f_i^{tree}\sim\int_{u_0}^1 du~\phi_\pi(u)T_H(u)~,
\label{LCSR}
\end{equation}
where $T_H$ is the process-dependent amplitude.
Here $u_0\equiv(m_b^2-q^2)/(s_0-q^2)$ where $s_0$ is the continuum threshold is
a lower limit of the convolution integral.
It scales as $u_0\sim 1-\Lambda_{\rm QCD}/m_b$;
thus only the highly asymmetric momentum configuration is relevant.
This is nothing but the exact meaning of the soft overlap.
Hard spectator interactions as well as the vertex corrections appear at
$\calO(\alpha_s)$.
Numerically Eq.\ (\ref{LCSR}) is dominant compared to the $\calO(\alpha_s)$
contributions.
\par
Recently, a new insight into the heavy-light transition was proposed
\cite{Lee:2006ew}.
It was worked out that there is an upper limit on the momentum of the outgoing 
quark from the weak vertex, less than the maximum recoil energy of $m_b/2$.
When the heavy quark is changed into the light quark with very high energy via
weak interaction, it suddenly moves through the ''brown muck'' consisting of
the light degrees of freedom.
The situation is very similar to the case when an electron goes through a dense
medium, where the Cherenkov radiation should occur.
Much more similar processes have been studied in the heavy-ion collisions 
recently.
Here an energetic parton enters through a dense hadronic medium, and possible
Cherenkov gluon radiation has been considered extensively
\cite{Koch:2005sx,Dremin:2005an}.
The energy loss due to the Cherenkov radiation can be easily calculated:
\begin{equation}
\frac{dE_c}{dx}=4\pi\alpha_s\int_{n(\epsilon)>1}d\epsilon~\epsilon
\left[1-\frac{1}{n^2(\epsilon)}\right]~,
\end{equation}
where $n(\epsilon)$ is the index of refraction.
The nonperturbative nature is encoded in $n(\epsilon)$.
The amount of energy loss for heavy-ion collisions varies around
$0.1\sim 1$ GeV/fm up to the model.
But it is generally accepted that the energy loss due to the Cherenkov 
radiation is smaller than the normal radiation by multiple scattering.
\par
Roughly speaking, the Cherenkov energy loss is about \cite{Koch:2005sx}
$dE_c/dx\sim 4\pi\alpha_s \ell_0^2/2$, where $\ell_0\sim\calO(\Lambda_{\rm QCD})$
is the gluon energy.
The total energy loss might be
\begin{equation}
\frac{E_c}{E}\sim 4\pi\alpha_s\frac{\ell_0^2L}{2E}
\sim\calO\left(\frac{\Lambda_{\rm QCD}}{m_B}\right)~,
\end{equation}
where $L\sim 1/\Lambda_{\rm QCD}$ is the flight length of the energetic parton
during the formation of $\pi$.
More precise estimation requires the detailed structure of the index of
refraction $n(\epsilon)$.
But this naive power counting is enough to give an important message for the
soft overlap.
If we take into account the Cherenkov energy loss, the convolution integral of
Eq.\ (\ref{LCSR}) will be changed into
\begin{equation}
f_i^{tree}\sim\int_{u_0}^{1-E_c/E}du~\phi_\pi(u)T_H(u)~.
\label{cutoff}
\end{equation}
Since $1-u_0\sim\calO(\Lambda_{\rm QCD}/m_B)\sim E_c/E$, the integration domain
shrinks severely.
Consequently the soft overlap is highly suppressed.
\par
It will be a good phenomenological trade to introduce the cutoff
$\ubar_c\equiv 1-u_c\equiv 1-E_c/E$ for the nonperturbative $n(\epsilon)$.
It was shown in \cite{Lee:2006ew} that for a
natural scale of $u_c\simeq m_\pi/m_B$ one gets a compatible
result for $B\to\pi$ form factors with other approaches.
\par
In this paper, we extend this approach to several $B\to P,V$ decays.
The extension is quite straightforward.
In the next Section, the relevant decay form factors are defined and the 
end-point singularities are identified.
In Sec.\ III, various form factors are evaluated with the momentum fraction
cutoffs, which originate from the Cherenkov energy loss.
Section IV contains discussions and conclusions.

\section{End-point singularities}
The form factors of $\Bbar$ decays into $P, V$ are defined by
\cite{Beneke:2000wa}
\begin{eqnarray}
\langle P(p)|\qbar\gamma^\mu b|\Bbar(p_B)\rangle
&=&
f_+(q^2)\left[p_b^\mu+p^\mu-\frac{m_B^2-m_P^2}{q^2}q^\mu\right]
+f_0(q^2)~\frac{m_B^2-m_P^2}{q^2}q^\mu~,\\
\langle P(p)|\qbar\sigma^{\mu\nu}q_\nu b|\Bbar(p_B)\rangle
&=&
\frac{if_T(q^2)}{m_B+m_P}\Big[q^2(p_B^\mu+p^\mu)-(m_B^2-m_P^2)q^\mu\Big]~,\\
\langle V(p,\epsilon^*)|\qbar\gamma^\mu b|\Bbar(p_B)\rangle
&=&
\frac{2iV(q^2)}{m_B+m_V}\epsilon^{\mu\nu\rho\sigma}
 \epsilon^*_\nu p_\rho p^B_\sigma~,\\
\langle V(p,\epsilon^*)|\qbar\gamma^\mu\gamma_5 b|\Bbar(p_B)\rangle
&=&
2m_V A_0(q^2)\frac{\epsilon^*\cdot q}{q^2}q^\mu
+(m_B+m_V)A_1(q^2)\left[\epsilon^{*\mu}-\frac{\epsilon^*\cdot q}{q^2}q^\mu\right]
\nn\\
&&
-A_2(q^2)\frac{\epsilon^*\cdot q}{m_B+m_V}\left[p_B^\mu+p^\mu-
 \frac{m_B^2-m_V^2}{q^2}q^\mu\right]~,\\
\langle V(p,\epsilon^*)|\qbar\sigma^{\mu\nu}q_\nu b|\Bbar(p_B)\rangle
&=&
2T_1(q^2)\epsilon^{\mu\nu\rho\sigma}\epsilon^*_\nu p^B_\rho p_\sigma~,\\
\langle V(p,\epsilon^*)|\qbar\sigma^{\mu\nu}\gamma_5 q_\nu b|\Bbar(p_B)\rangle
&=&
-iT_2(q^2)\Big[(m_B^2-m_V^2)\epsilon^{*\mu}-(\epsilon^*\cdot q)(p_B^\mu+p^\mu)
\Big]\nn\\
&&
-iT_3(q^2)(\epsilon^*\cdot q)\left[q^\mu-\frac{q^2}{m_B^2-m_V^2}(p_B^\mu+p^\mu)
\right]~,
\end{eqnarray}
where
\begin{eqnarray}
\epsilon^{0123}&=&-1~,\\
q^\mu&\equiv&p_B^\mu-p^\mu~,\\
\epsilon^\mu&:& {\rm polarization~vector~of~}V~,\\
m_M&:& {\rm mass~of~}M=B,~P,~V~.
\end{eqnarray}

\par
\begin{figure}
\includegraphics{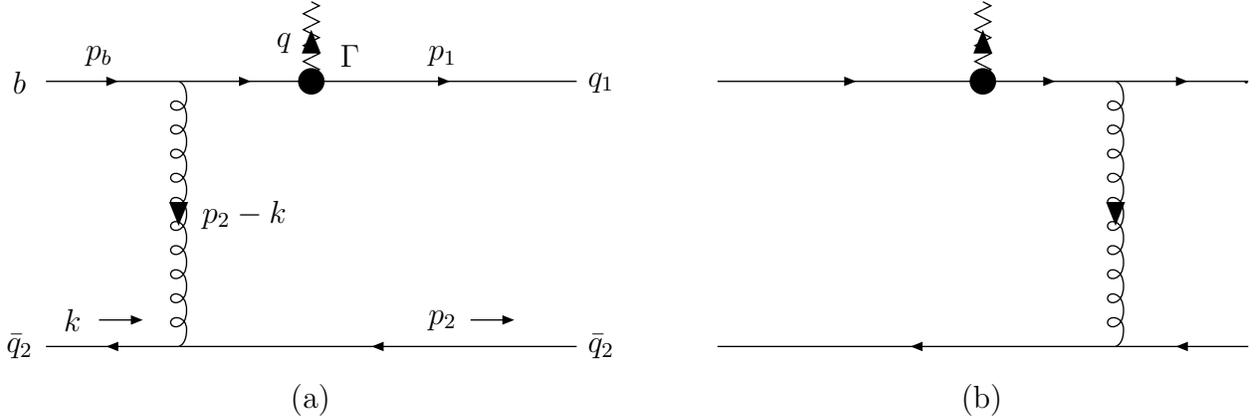}\\
\caption{\label{diagram}Gluon exchanges with the spectator quarks.}
\end{figure}

The end-point singularity appears when one considers the gluon exchange diagrams
with the spectator quarks (Fig.\ \ref{diagram}).
Their amplitude is proportional to the scattering kernel \cite{Beneke:2000wa}
\begin{equation}
T_{ij\ell m}=
-g^2C_F\frac{\gamma^\mu_{\ell m}}{(p_2-k)^2}
\left[
\Gamma~\frac{\pslash_b+\kslash-\pslash_2+m_b}{(p_b+k-p_2)^2-m_b^2}~\gamma_\mu
+\gamma_\mu~\frac{\pslash_1-\kslash+\pslash_2}{(p_1-k+p_2)^2}~\Gamma
\right]_{ij}~,
\end{equation}
which will be convoluted with the meson DAs to give the full amplitude.
At the heavy quark and large recoil limit, the kernel reduces to
\begin{equation}
T_{ij\ell m}\simeq
-g^2C_F\gamma^\mu_{\ell m}\left[
\Gamma~\frac{m_b(1+\vslash)-\ubar E\nslash}{4\ubar^2m_bE^2k^+}~\gamma_\mu
+\gamma_\mu~\frac{E\nslash-\kslash}{4\ubar E^2(k^+)^2}~\Gamma\right]_{ij}~,
\label{kernel}
\end{equation}
where $p_b=m_b v$, $p=En$, $p_1=up$, $p_2=\ubar p\equiv(1-u)p$, and
$k^+=n\cdot k$ with $n$ being a light-like vector.
The end-point singularity occurs when $\ubar\to 0$ or $k^+\to 0$ unless the
meson DAs fall fast enough to compensate the factors of $\ubar^n$ or 
$(k^+)^n$ in the denominator.
Actually, the asymptotic form of DAs of the light mesons are 
$\phi^{\rm ays}(u)\sim u\ubar$.
As for $B$ DAs, the specific forms depend on models, but they can also suffer
from the end-point singularity with $1/(k^+)^2$ factors.
In this work, we just concentrate on the problem of $\ubar\to 0$ for simplicity,
because it is conceptually more important.
\par
From this observation, it is commonly argued that heavy-light transition is
dominated by the ''soft'' physics where $\ubar\to 0$.
And the troublesome divergent terms are absorbed into the nonperturbative
''soft'' form factor \cite{Beneke:2000wa}.
Dominance of the soft-overlap contribution is then supported by the 
phenomenological fitting within this parametrization.
Note that near the end point where $\ubar\to 0$, $p_1\to p$ and $p_2\to 0$;
the parton momenta of the outgoing light meson are highly asymmetric.
However, one cannot conclude that the soft physics dominates from unphysical
divergences.
Even worse is that we cannot extrapolate the nonperturbative physics from the
perturbative analysis.
When $\ubar$ is very close to 0, the offshellness of the exchanged gluon
momentum $(p_2-k)^2\sim -2\ubar Ek^+$ is vanishing, so we are entering the
nonperturbative regime.
In this case the neglected terms of order $\sim\calO(\Lambda_{\rm QCD})$
become significant, and thus Eq.\ (\ref{kernel}) is not reliable any longer.
\par
At this stage, it will be very interesting to see how the ''soft'' form factor
is treated in SCET.
In SCET the soft form factor is defined by a series of operators containing
{\em collinear} quarks only \cite{Bauer:2002aj}.
In other words, energetic final-state meson is described solely by the collinear
quarks; soft+collinear combination is not allowed.
This is also justified by the large rapidity gap in \cite{Manohar:2006nz}.
That's the reason why the defining operators have the interaction Lagrangians
$\calL_{\xi q}$ or messenger modes which convert the soft spectator quark ($q$)
into collinear one ($\xi$) through the exchange of collinear gluons or into the 
soft-collinear quark.
In this sense, there is no soft-overlap contributions in SCET {\em a priori.}
\par
This makes a sharp contrast with the works of 
\cite{Lange:2003pk,Ball:2003bf} where the soft-overlap contribution plays an 
important role.
Also in many literatures the end-point singular terms are absorbed into the
soft form factor as in Eq.\ (\ref{soft}) \cite{Beneke:2000wa,Beneke:2003pa}.
Thus the ''soft'' form factor $\xi_i^M$ contains both soft overlap and hard
spectator interactions.
This parametrization is not bad in a viewpoint of the spin symmetry because
all the terms in the soft form factor satisfy the spin-symmetry relations.
But it may cause some confusions; for example, it is not adequate to directly
compare $\xi_i^M$ with $\zeta^{BM}$ since the soft overlap is included in the
former while not in the latter from the construction.
As pointed out in \cite{Beneke:2003pa}, it might be unfruitful to extract the
hard scattering effects from $\xi_i^M$ to leave it purely nonperturbative.
However, it is at least conceptually important to separate the soft overlap
from the hard scattering when which of the two is dominant matters.
\par
One more profound matter in SCET regarding the end-point singularity is double
counting.
The phase space region where $\ubar\to 0$ corresponds to the zero-bin of the
collinear momentum, which must be subtracted to avoid double counting
\cite{Manohar:2006nz}.
The zero-bin subtraction is equivalent to introducing the IR cutoff regulator.
\par
As argued in Sec.\ I, there is a new possibility for the energy cutoff due to
the Cherenkov gluon radiation in the full theory.
The origin of this cutoff is of course quite different from that of zero-bin
subtraction;
the latter is necessary for an effective theory where the various momentum
scales are involved.
It is related to the self-consistency of the effective theory.
But there is a situation where only collinear modes are considered, as in 
producing energetic light mesons from heavy-light decays.
In this case the zero-bin subtraction corresponds to establishing a cutoff
below which one should not enter.
The Cherenkov energy loss explains why this must be the case;
the energetic parton cannot reach far near end point.
\par
The suppression of soft overlap can be easily seen in 
Eqs.\ (\ref{LCSR}), (\ref{cutoff}).
In numerical calculations, however, the value of $u_0$ is not so close to 1.
Typically, $u_0\approx 0.65\sim 0.70$ for $B\to\pi$.
Its deviation from unity is much larger than the usual
$\Lambda_{\rm QCD}/m_B\approx0.04$ for $\Lambda_{\rm QCD}\sim 200$ MeV.
It is very difficult and ambiguous to determine what portion of momentum should
be transferred to insure the soft overlap configuration, or to make soft quark
collinear.
But it is quite true that Eq.\ (\ref{LCSR}) contains more than ''soft overlap''
with $u_0\approx 0.65\sim 0.70$ in numerics.
Furthermore, the approximation in \cite{Ball:2003bf}
\begin{equation}
f_+^{tree}\sim\int_{u_0}^1du~\phi_\pi(u)
\approx-\frac{1}{2}\phi'_\pi(1)\ubar_0^2\simeq 0.35~,
\end{equation}
tends to increase the numerical value compared to the original integral
($\simeq 0.27$).
This is because $\ubar_0\equiv 1-u_0\simeq 0.34$ is not sufficiently small.
In short, the soft overlap contribution is overestimated in LCSR.
The pure soft overlap contribution comes from the much narrower range of
momentum fraction, which would be shrunken again by the Cherenkov radiation.

\section{Form factors with momentum fraction cutoffs}

In what follows, we {\em assume} that the soft overlap is negligible because
of the cutoff by the Cherenkov energy loss.
This approach is on the same line as pQCD or SCET where the soft overlap is
ignored.
The $B\to P,V$ form factors are then given by the hard gluon exchange 
processes. 
There is now one nonperturbative parameter $u_c$ which regulates the
divergent convolution as a cutoff.
Explicitly \cite{Beneke:2000wa},
\begin{eqnarray}
f_+&=&\left(\frac{\alpha_sC_F}{4\pi}\right)
\left(\frac{\pi^2f_Bf_P m_B}{N_cE^2}\right)
\int_{u_c}^{\ubar_c}du\int_0^{\infty}dk^+\left\{
\frac{4E-m_b}{m_b}~\frac{\phi(u)\phi^B_+(k^+)}{\ubar k^+}\right.\nn\\
&&
\left.
+\frac{1+\ubar}{\ubar^2 k^+}\phi(u)\phi^B_-(k^+)
+\frac{\mu_\pi}{2E}\left[\frac{1}{
 \ubar^2 k^+}\left(\phi_p(u)-\frac{\phi'_\sigma(u)}{6}\right)
 +\frac{4E}{\ubar (k^+)^2}\phi_p(u)\right]\phi^B_+(k^+)\right\}~,\\
f_0&=&\left(\frac{2E}{m_B}\right)f_+
 +\left(\frac{\alpha_sC_F}{4\pi}\right)\left(\frac{m_B-2E}{2E}\right)
 \Delta F_P~,\\
f_T&=&\left(\frac{m_B+m_P}{m_B}\right)f_+
 -\left(\frac{\alpha_sC_F}{4\pi}\right)\left(\frac{m_B+m_P}{2E}\right)
 \Delta F_P~,
\end{eqnarray}
for $B\to P$, where $N_c=3$, $\mu_\pi=m_P^2/(m_1+m_2)$, and 
$\phi_{p,\sigma}(u)$ are the higher twist DAs.
Here,
\begin{equation}
\Delta F_P=
\left(\frac{8\pi^2f_Bf_P}{N_cm_B}\right)
\langle k_+^{-1}\rangle_+\langle\ubar^{-1}\rangle_P~,
\end{equation}
where
\begin{eqnarray}
\langle k_+^{-1}\rangle_+&\equiv&\int dk^+~\frac{\phi_+^B(k^+)}{k^+}~,\nn\\
\langle\ubar^{-1}\rangle_P&\equiv&\int du~\frac{\phi(u)}{\ubar}~.
\end{eqnarray}
\par
For $B\to V$, we have
\begin{eqnarray}
V&=&
\left(\frac{\alpha_sC_F}{4\pi}\right)\left(\frac{\pi^2f_Bm_B}{N_cE^2}\right)
\left(\frac{m_B+m_V}{m_b}\right)\int_{u_c}^{\ubar_c}du\int_0^\infty dk^+\Bigg\{
\nn\\
&&
\frac{f_\perp\phi_\perp(u)\phi_-^B(k^+)}{\ubar^2 k^+}
+\frac{f_Vm_V}{2E}\left(\frac{2E}{\ubar (k^+)^2}+\frac{1}{\ubar^2 k^+}\right)
\left[g^{(v)}_\perp(u)-\frac{g'^{(a)}_\perp(u)}{4}\right]\phi_+^B(k^+)\Bigg\}~,
\\
A_0&=&
\left(\frac{\alpha_sC_F}{4\pi}\right)\left(\frac{\pi^2f_Bm_B}{N_cE^2}\right)
\int_{u_c}^{\ubar_c}du\int_0^\infty dk^+\Bigg\{
\frac{f_V\phi_\parallel(u)\phi_+^B(k^+)}{\ubar k^+}
+\frac{1+\ubar}{\ubar^2 k^+}f_V\phi_\parallel(u)\phi_-^B(k^+)\nn\\
&&
+\frac{f_\perp m_V}{2E}\left[
-\frac{2E}{\ubar(k^+)^2}h'^{(s)}_\parallel(u)\phi_+^B(k^+)
+\frac{1}{\ubar^2k^+}\left(h^{(t)}_\parallel(u)-\frac{h'^{(s)}_\parallel}{2}\right)
\phi_+^B(k^+)\right]\Bigg\}~,\\
A_1&=&\left[\frac{2Em_B}{(m_B+m_V)^2}\right]V~,\\
A_2&=&
V-\left[\frac{m_V(m_B+m_V)}{Em_B}\right]A_0
+\left(\frac{\alpha_sC_F}{4\pi}\right)\frac{m_V(m_B+m_V)}{Em_B}
\frac{m_B^2-2m_bE}{4E^2}\Delta F_\parallel~,\\
T_1&=&
\left(\frac{m_B}{m_B+m_V}\right)V
+\left(\frac{\alpha_sC_F}{4\pi}\right)\frac{m_B}{4E}\Delta F_\perp~,\\
T_2&=&
\left(\frac{2E}{m_B+m_V}\right)V
+\left(\frac{\alpha_sC_F}{4\pi}\right)\frac{1}{2}\Delta F_\perp~\\
T_3&=&
\left(\frac{m_B}{m_B+m_V}\right)V-\left(\frac{m_V}{E}\right)A_0
+\left(\frac{\alpha_sC_F}{4\pi}\right)
\left(\frac{m_B}{4E}\Delta F_\perp+\frac{m_Vm_B^2}{4E^3}\Delta F_\parallel\right)~.
\end{eqnarray}
As in the $B\to P$ case, we define
\begin{eqnarray}
\Delta F_\perp&=&
\left(\frac{8\pi^2 f_Bf_\perp}{N_cm_B}\right)
\langle k_+^{-1}\rangle_+\langle\ubar^{-1}\rangle_\perp~,\\
\Delta F_\parallel&=&
\left(\frac{8\pi^2 f_Bf_V}{N_cm_B}\right)
\langle k_+^{-1}\rangle_+\langle\ubar^{-1}\rangle_\parallel~,
\end{eqnarray}
where
\begin{equation}
\langle\ubar^{-1}\rangle_{\perp,\parallel}\equiv
\int du~\frac{\phi_{\perp,\parallel}(u)}{\ubar}~.
\end{equation}
The twist-3 DAs $g_\perp^{(v,a)}(u)$ and $h_\parallel^{(t,s)}(u)$ can be 
written in terms of the twist-2 DAs by Wandzura-Wilczek type relations
\cite{Ball:1998sk,Beneke:2000wa}
\begin{eqnarray}
g_\perp^{(v)}(u)&=&
\frac{1}{2}\left[\int_0^u\frac{\phi_\parallel(v)}{\vbar}~dv
+\int_u^1\frac{\phi_\parallel(v)}{v}~dv\right]+\cdots~,\nn\\
g_\perp^{(a)}(u)&=&
2\left[\ubar\int_0^u\frac{\phi_\parallel(v)}{\vbar}~dv
+u\int_u^1\frac{\phi_\parallel(v)}{v}~dv\right]+\cdots~,\nn\\
h_\parallel^{(t)}(u)&=&
(2u-1)\left[\int_0^u\frac{\phi_\perp(v)}{\vbar}~dv
-\int_u^1\frac{\phi_\perp(v)}{v}~dv\right]+\cdots~,\nn\\
h_\parallel^{(s)}(u)&=&
2\left[\ubar\int_0^u\frac{\phi_\perp(v)}{\vbar}~dv
+u\int_u^1\frac{\phi_\perp(v)}{v}~dv\right]+\cdots~.\nn\\
\end{eqnarray}
Note that the integration domain is changed into
$\int_0^1du\to\int_{u_c}^{\ubar_c}du$ to avoid the soft overlap region.
The integral over $k^+$ will cause another divergence at $k^+=0$.
We simply introduce an IR cutoff $\bar\Lambda=m_B-m_b$ for $B$ meson sector.
To get the numerical estimate, we use the asymptotic form of
$\phi$, $\phi_{p,\sigma,\parallel,\perp}$, and
\begin{equation}
\phi^B_+(\omega)=\frac{\omega}{\omega_0^2}e^{-\omega/\omega_0}~,~~~
\phi^B_-(\omega)=\frac{1}{\omega_0}e^{-\omega/\omega_0}~,
\label{phiB}
\end{equation}
where $\omega_0$ is a model parameter \cite{Grozin:1996pq}.
The results are summarized in Tables \ref{B2P}, \ref{B2V} for some $u_c$s.
\begin{table}
\begin{tabular}{c|cccc}
$u_c$ &~~~$\frac{m_\pi}{m_B}$ & $0.02$ & 0.0213 & LCSR \cite{Ball:2001fp}\\\hline
$f_+^{\pi}(0)$ & ~~~0.212 & 0.273& 0.258 & 0.258\\
$f_T^{\pi}(0)$ &~~~ 0.185 & 0.247& 0.231 & 0.253\\
$f_+^{K}(0)$ &~~~ 0.265 & 0.341& 0.322 & 0.331\\
$f_T^{K}(0)$ &~~~ 0.248 & 0.330& 0.309 & 0.358\\
$f_+^{\eta}(0)$ &~~~ 0.221 & 0.285& 0.269 & 0.275\\
$f_T^{\eta}(0)$ &~~~ 0.209 & 0.279& 0.262 & 0.285
\end{tabular}
\caption{\label{B2P} $B\to P$ form factors for various $u_c$.
$\alpha_s$ is taken at $\mu=\sqrt{m_B\Lambda_{\rm QCD}}\simeq1.47$ GeV.}
\end{table}

\begin{table}
\begin{tabular}{c|cccc}
$u_c$ &~~~$\frac{m_\pi}{m_B}$ & $0.02$ & 0.0135 & LCSR \cite{Ball:2004rg}\\\hline
$V^{\rho}(0)$ & ~~~0.162 & 0.217 & 0.324 & 0.323\\
$A_0^{\rho}(0)$ & ~~~0.225 & 0.298 & 0.441 & 0.303\\
$A_1^{\rho}(0)$ & ~~~0.123 & 0.165& 0.247 & 0.242\\
$A_2^{\rho}(0)$ & ~~~0.089 & 0.119 & 0.178 & 0.221\\
$T_1^{\rho}(0)$ & ~~~0.161 & 0.209 & 0.303 & 0.267\\
$T_3^{\rho}(0)$ & ~~~0.110 & 0.137 & 0.189 & 0.176\\\hline
$V^{\omega}(0)$ & ~~~0.156 & 0.209 & 0.313 & 0.293\\
$A_0^{\omega}(0)$ & ~~~0.208 & 0.275 & 0.407 & 0.281\\
$A_1^{\omega}(0)$ & ~~~0.118 & 0.159 & 0.237 & 0.219\\
$A_2^{\omega}(0)$ & ~~~0.087 & 0.117 & 0.176 & 0.198\\
$T_1^{\omega}(0)$ & ~~~0.154 & 0.200 & 0.290 & 0.242\\
$T_3^{\omega}(0)$ & ~~~0.106 & 0.133 & 0.184 & 0.155\\\hline
$V^{K^*}(0)$ & ~~~0.201  & 0.269 & 0.403 & 0.411\\
$A_0^{K^*}(0)$ & ~~~0.296 & 0.359 & 0.535 & 0.374\\
$A_1^{K^*}(0)$ & ~~~0.147 & 0.197 & 0.295 & 0.292\\
$A_2^{K^*}(0)$ & ~~~0.096 & 0.129 & 0.193 & 0.259\\
$T_1^{K^*}(0)$ & ~~~0.192 & 0.251 & 0.366 & 0.333\\
$T_3^{K^*}(0)$ & ~~~0.119 & 0.148 & 0.203 & 0.202\\
\end{tabular}
\caption{\label{B2V} $B\to V$ form factors for various $u_c$.}
\end{table}
\par
In these Tables, $u_c=m_\pi/m_B$ is given as a natural representative
scale in heavy-light decays.
Note that the larger $u_c$, the smaller form factors.
Results from LCSR is also listed for comparison.
The value of $u_c=0.0213~(0.0135)$ is chosen to coincide the value of 
$f_+^\pi(0)$ ($V^\rho(0)$) with that from LCSR.
Comparing the columns of $u_c=0.0213~(0.0135)$ and LCSR, $f_i$s for $B\to P$
are generally in good agreement except $f_T^K(0)$.
As for $B\to V$, we can find the following tendency; 
$A_0^{\rm cutoff}\gg A_0^{\rm LCSR}$, 
$A_1^{\rm cutoff}\sim A_1^{\rm LCSR}$,
$A_2^{\rm cutoff}\lesssim A_2^{\rm LCSR}$,
$T_1^{\rm cutoff}> T_1^{\rm LCSR}$,
$T_3^{\rm cutoff}\sim T_3^{\rm LCSR}$.
\section{Discussions and conclusions}

Present analysis is done with the asymptotic forms of light meson DAs.
Typically, nonasymptotic effects contribute by a few tens of percent, 
so inclusion of them will be required to refine the results.
One of the main source of uncertainty is the $B$ meson DA, 
$\phi_\pm^B(\omega)$.
Usually the problematic divergent terms involving $\phi_\pm^B$ are also
absorbed into the soft form factors.
The remaining part contains the inverse moment of $\phi_+^B$;
\begin{equation}
\int_0^\infty d\omega~\frac{\phi_+^B(\omega)}{\omega}\equiv\frac{1}{\lambda_B}~.
\label{lambdaB}
\end{equation}
With Eq.\ (\ref{phiB}), one simply has $\lambda_B=\omega_0$.
We fix $\omega_0=2{\bar\Lambda}/3=0.32$ GeV \cite{Grozin:1996pq}.
The divergent moment is parametrized as
\begin{equation}
\frac{1}{\kappa_B}\equiv
\int_0^\infty d\omega~\frac{\phi_-^B(\omega)}{\omega}
\to\int_{\bar\Lambda}^\infty d\omega~\frac{\phi_-^B(\omega)}{\omega}~,
\end{equation}
where $\bar\Lambda$ acts as an IR regulator.
In this parametrization with Eq.\ (\ref{phiB}), the singular inverse-square 
moment over $\phi_+^B$ is
\begin{equation}
\int_0^\infty d\omega~\frac{\phi_+^B(\omega)}{\omega^2}
=\frac{1}{\omega_0\kappa_B}~.
\end{equation}
\par
It might be that the bad thing is not the singularity itself;
the singular {\em behavior} would be even worse in the sense that the form 
factors are very sensitive to the cutoff or regulator.
At least, we need to reduce the hadronic uncertainty on, say, $\lambda_B$.
But current analysis shows that for a reasonable set of cutoffs the form factors
are compatible with the literature.
The point is that there must be a fundamental cutoff in heavy-light decay due
to the Cherenkov energy loss.
\par
The necessary condition for the Cherenkov radiation is 
${\rm Re}[n(\epsilon)]>1$,
where $n(\epsilon)$ is the index of refraction.
Analogous to the photon case, $n(\epsilon)-1$ is proportional to the
forward scattering amplitudes $F(\epsilon)$.
At low energies, ${\rm Re}[F(\epsilon)]>0$ if $\epsilon>\epsilon_R$
for the Breit-Wigner resonance
$F(\epsilon)\sim(\epsilon-\epsilon_R+i\Gamma/2)^{-1}$ where $\epsilon_R$ is the
resonant energy and $\Gamma$ is the decay width \cite{Dremin:2005an}.
Since the light mesons are possible intermediate resonances of the brown muck,
the necessary condition can be easily satisfied also in $B\to P,V$ transitions.
\par
In exclusive decays, the Cherenkov gluons will eventually couple and transfer
the energy to the light degrees of freedom to make the final state meson.
On the other hand, in inclusive decays, they can appear as conelike jet events.
If the jets form a specific angle which cannot be explained by conventional
arguments, then they will be a strong candidate for the Cherenkov gluons.
\par
At a glance, the suppression of soft overlap by Cherenkov energy loss looks 
like Sudakov effects.
But the Cherenkov radiation is purely medium effects.
The Sudakov factor plays a role of weight for meson DAs which falls very
rapidly (though controversial) near the end-point region.
However, this is basically the extrapolation into the nonperturbative regime 
from the perturbative calculations.
The spirit of this work is that we should not {\em imagine} the nonperturbative
physics in the line of perturbative approaches.
The index of refraction $n(\epsilon)$ is a robust nonperturbative object which
forbids the energetic parton from the weak vertex to go far into the end-point
region.
\par
In conclusion, we propose a fundamental cutoff for the
heavy-to-light transitions due to possible Cherenkov gluon radiation.
It naturally avoids the end-point singularity and double counting problems.
In this picture the soft overlap contribution is highly suppressed; 
heavy-to-light decay is dominated by the hard scattering processes.
Though the numerical values of the weak form factor are very sensitive to the 
choice of the cutoff, for a natural scale of $u_c$ one gets a compatible
result with other approaches.
The existence of Cherenkov gluons can be directly checked from the conelike
jets in inclusive decays.

\begin{acknowledgments}
The author thanks Hsiang-nan Li for helpful discussions.
\end{acknowledgments}

\end{document}